\documentstyle[11pt,newpasp,twoside]{article}
\input epsf.tex 

\def\edcomment#1{\iffalse\marginpar{\raggedright\sl#1\/}\else\relax\fi}
\reversemarginpar

\input epsf.tex 

\newcommand{\mincir}{\raise -2.truept\hbox{\rlap{\hbox{$\sim$}}\raise5.truept 
\hbox{$?$}\ }} 
\newcommand{\gr}{\kern 2pt\hbox{}^\circ{\kern -2pt K}} 
\newcommand{\magcir}{\raise -2.truept\hbox{\rlap{\hbox{$\sim$}}\raise5.truept 
\hbox{$?$}\ }}

\newcommand{\be}{\begin{equation}} 
\newcommand{\ee}{\end{equation}} 
\newcommand{\bea}{\begin{eqnarray}} 
\newcommand{\eea}{\end{eqnarray}}

\newcommand{\etal}{{et al.}} 

\begin{document}

\title{
Constraints on the tensor mode from \\
large scale structure observations
}

\author{B.~Novosyadlyj$^{*}$, R.~Durrer$^{\dagger}$ and S.~Apunevych$^{*}$}

\affil{$^*$Astronomical Observatory of L'viv National University, Ukraine}
\affil{$^{\dagger}$ Department de Physique Th\'eorique, Universit\'e de Gen\`eve, Switzerland\\
}

\begin{abstract}
\noindent Observational data on the large scale structure (LSS) of the Universe 
are used to establish an upper limit for the amplitude of the tensor mode 
marginalized over all other cosmological parameters within the class of adiabatic 
inflationary models. It is shown that the upper 1$\sigma$ limit for the contribution 
of a tensor mode to the COBE DMR data is T/S$<1$. 
\end{abstract}  

The ratio of tensor to scalar fluctuations (T/S) is important
for discriminating among large number of inflation models as well
as for the determination of the cosmological parameters.
 The tensor mode contributes to the temperature fluctuations
of cosmic microwave background (CMB) at the largest angular scales. 
Its amplitude  can be determined by comparing the 4-year COBE data 
(Bennett \etal~1996)  with the large scale structure observations to
which only the scalar mode contributes. 
The main reason why present data does not lead to a stringent limit is that 
the inferred contribution of a tensor mode strongly depends on
practically all other cosmological parameters. Most notable on the
spectral index of primordial power spectrum of scalar perturbations,
$n_s$, but also on the values of cosmological constant $\Omega_\Lambda$, 
the dark matter content and its nature, 
spatial curvature $\Omega_k$, the Hubble parameter h
and the baryon content $\Omega_b$. This means that 
T/S must be determined simultaneously with all these parameters
from wide range scale of cosmological observations. 
Similar investigations show
that mixed dark matter model with cosmological constant ($\Lambda$MDM) 
can explain virtually all cosmological measurements (Novosyadlyj et al. 2000).  
The goal of this paper is to determine the upper limit on the ratio T/S 
in the framework of $\Lambda$MDM models. We restrict ourselves 
the sub-class of models without early reionization.

We find the parameters of cosmological model which matches the 
observational data on the large scale structure of the Universe best and by
marginalization over all other parameters determine the upper limit
on T/S. 

\medskip

\medskip
 
Our approach is based on the comparison of observational data on the 
structure of the Universe over a wide range of scales with theoretical 
predictions from the linear power spectrum of  density fluctuations.
The form of the spectrum strongly depends on the 
cosmological parameters $\Omega_m$, $\Omega_b$, $\Omega_{\nu}$, $N_{\nu}$,
$h$ and $n_s$. Minimization of the quadratic differences 
between the theoretical and observational values divided by the observational 
errors, $\chi^2$, determines the best-fit values for the 
above mentioned cosmological parameters and the amplitude of the power spectrum 
of scalar mode. For this we use the following
observational data set: 
the location and amplitude of the first acoustic  peak in the angular power 
spectrum of the CMB temperature fluctuations deduced from the CMB map obtained 
in the Boomerang and MAXIMA-1 experiments; the power spectrum of density 
fluctuations of Abell-ACO clusters obtained from their space distribution;
the constraint for the amplitude of the fluctuation power spectrum on 
$\approx 10h^{-1}$Mpc scale derived from a recent optical determination of the 
mass function of nearby galaxy clusters, from the  evolution of the galaxy 
cluster X-ray temperature distribution function and from the existence of three 
very massive clusters of galaxies observed so far at $z>0.5$; 
bulk flows of galaxies in sphere of radius 
$50h^{-1}$Mpc; 
the constraint for the amplitude of the fluctuation power spectrum on 
$0.1-1h^{-1}$Mpc scales and $z\approx2-3$ derived from the Ly-$\alpha$ 
absorption lines seen in quasar spectra; the data on the direct measurements 
of the Hubble constant $\tilde h=0.65\pm 0.10$ which is a compromise between 
results obtained by different groups; the nucleosynthesis constraint
on the baryon density derived from the abundance   of inter galactic deuterium 
$\widetilde{\Omega_bh^2} = 0.019\pm 0.0024$. (For references and more
details of the observations used see
Durrer \& Novosyadlyj 2000).

\smallskip
 
One of the main ingredients for the solution for our search problem 
is a reasonably fast and accurate determination of the  linear 
transfer function for dark matter clustering 
which depends on the cosmological parameters. We use accurate analytical 
approximations of the MDM transfer function $T(k;z)$ depending on 
the parameters $\Omega_m$, $\Omega_b$, $\Omega_{\nu}$, $N_{\nu}$ and 
$h$ given by Eisenstein \& Hu (1999). The 
linear power spectrum of matter density fluctuations
$P(k;z)=A_sk^{n_s}T^2(k;z)D_1^2(z)/D_1^2(0)$, 
where $A_s=2\pi^{2}\delta_{h}^{2}(3000{\rm Mpc}/h)^{3+n_s}$
is the normalization constant for scalar perturbations, 
$D_1(z)$ is the linear growth factor and $\delta_{h}$ is the matter 
density fluctuation at horizon scale.

The Abell-ACO power spectrum is related to the matter power 
spectrum at $z=0$, $P(k;0)$, by the cluster biasing parameter $b_{cl}$:
$P_{A+ACO}(k)=b_{cl}^{2}P(k;0)$.
We assume scale-independent linear bias as free parameter of which the
best-fit value is determined together with the other cosmological parameters.
 
The dependence of the position and amplitude of the first acoustic 
peak in the CMB power spectrum on cosmological 
parameters $n_s$, $h$, $\Omega_b$, $\Omega_{cdm}$ and 
$\Omega_{\Lambda}$ is obtained with the analytical approximation 
given by Efstathiou \& Bond (1999) which has been  extended to  models
with non-zero  
curvature  ($\Omega_k\equiv1-\Omega_m-\Omega_{\Lambda}\ne0$) by
Durrer \& Novosyadlyj (2000). The accuracy of this approximation in
the parameter ranges which we consider is better then 5\%.

The theoretical values of the other experimental constraints are
calculated as described in Durrer \& Novosyadlyj (2000). 

\medskip 
We consider the normalization  of the scalar mode $\delta_h^{LSS}$ 
as free parameter which is determined together with the cosmological parameters 
$\Omega_m$, $\Omega_{\Lambda}$, $\Omega_{\nu}$, $N_{\nu}$, $\Omega_b$,
$h$, $n_s$ and $b_{cl}$ by the Levenberg-Marquardt $\chi^2$  
minimization method. Hence, we have eight free parameters 
(the number species of massive neutrino is discrete and fixed). 
The formal number of observational points is 24 but, as it was shown
in Novosyadlyj et al. (2000), the 13 points of the cluster power
spectrum  can be described by just 3 degrees of freedom, so that the  
maximal number of truly independent measurements is 14. Therefore,  
the number of degrees of freedom for our search procedure is $N_F=
N_{\rm exp}-N_{\rm par}= 6$. The  
model with one sort of massive neutrinos provides the best fit to  
the data, $\chi^2_{min}=5.8$. The best fit parameters are 
$\delta_h^{LSS}=(2.95\pm2.55)\cdot 10^{-5}$,
$\Omega_m=0.40\pm0.08$, $\Omega_{\Lambda}=0.66\pm0.07$, 
$\Omega_{\nu}=0.05\pm0.05$, $\Omega_b=0.038\pm0.010$, $n_s=1.14\pm0.31$, $h=0.71\pm0.09$  
and $b_{cl}=2.4\pm0.3$ (standard errors). 
The best-fit value for density perturbation at horizon 
scale from the 4-year COBE data $\delta_h^{COBE}$ for the same model is 
larger then the best-fit value determined from 
LSS characteristics, $\delta_h^{COBE}=4.0\cdot10^{-5}>\delta_h^{LSS}$. This 
implies that COBE $\Delta T/T$ data may contain a non-negligible tensor 
contribution. The ratio $T/S$ is given by  
$T/S=(\delta_h^{COBE}-\delta_h^{LSS})/\delta_h^{LSS}$. The best fit
value of this parameter implied by the 
best-fit values of $\delta_h^{COBE}$ and $\delta_h^{LSS}$ from the
data used above is  $T/S=0.36$, if we use just
the Boomerang data (de Bernardis \etal~2000) for the amplitude and
position of the first acoustic peak,  and $T/S=0.18$  
from the combined Boomerang + MAXIMA-1 (Hu \etal~2000) data. Since the
standard error is  
rather large, $\approx90\%$, we determine upper confidence limits for $T/S$
by marginalizing $\delta_h^{LSS}$ over all the other parameters. This
procedure yields 
$T/S<1$ at 1$\sigma$ C.L. and $T/S<1.5$ at 2$\sigma$ C.L. from the 
Boomerang data alone for the  amplitude and position of the first 
acoustic peak. If we use the combined Boomerang + MAXIMA-1 data  
the limits are somewhat lower, 0.9 and 1.3 correspondingly, due to 
the higher amplitude of the first acoustic peak measured by MAXIMA-1. 
The 1$\sigma$ upper constraint on the tensor mode obtained recently by 
Kinney \etal~ (2000) from the Boomerang and MAXIMA-1 data on the CMB 
power spectrum for  the same class  of models (T/S$<0.8$ in our definition)
is close to the value obtained here.

\medskip

{\it Acknowledgments:}

B. Novosyadlyj is grateful to SOC for an IAU travel grant for the 
participation in IAU Symposium 201.

\end{document}